\documentclass[
letter,
twocolumn,
superscriptaddress,
amsmath,
amssymb,
prl,
nofootinbib
showkeys,
10pt,
floatfix,
nobibnotes,
aps,
fltpage
]{revtex4-2}

\usepackage{latexsym}
\usepackage{amsmath}
\usepackage{amssymb}
\usepackage[T1]{fontenc}
\usepackage[open]{bookmark}
\usepackage{hyperref}
\hypersetup{colorlinks=true,allcolors=blue}
\usepackage{upgreek}

\usepackage{orcidlink}

\usepackage{dcolumn}
\usepackage{bm}
\usepackage{xcolor}
\usepackage{hyperref}
\usepackage{cleveref}
\usepackage{float}
\usepackage{svg}
\usepackage{amssymb}
\usepackage{scalerel}
\usepackage{caption}
\usepackage{wasysym}
\usepackage{lipsum}
\usepackage{enumitem}

\providecommand{\pnl}[1]{{\textcolor{black}{(#1)}}}

\makeatletter
\pretocmd\frontmatter@keys@format{\addvspace{20\p@}}{}{}
\makeatother

\usepackage{svg}
\usepackage{cleveref}
\usepackage{siunitx}
\usepackage{caption}
\usepackage{multirow}
\usepackage{makecell}
\usepackage{tabularray}
\usepackage{makecell} 
\usepackage{array} 
\usepackage{graphicx} 
\usepackage{booktabs} 

\begin{document}

\title{Substrate-Assisted Cathodoluminescence}

\author{Sven Ebel\,\orcidlink{0009-0005-3224-6413}}
\affiliation{POLIMA---Center for Polariton-driven Light--Matter Interactions, University of Southern Denmark, Campusvej 55, DK-5230 Odense M, Denmark}

\author{N. Asger Mortensen\,\orcidlink{0000-0001-7936-6264}}
\affiliation{POLIMA---Center for Polariton-driven Light--Matter Interactions, University of Southern Denmark, Campusvej 55, DK-5230 Odense M, Denmark}
\affiliation{Danish Institute for Advanced Study, University of Southern Denmark, Campusvej 55, DK-5230 Odense M, Denmark}

\author{Sergii Morozov\,\orcidlink{0000-0002-5415-326X}}
\email{Corresponding author: semo@mci.sdu.dk}
\affiliation{POLIMA---Center for Polariton-driven Light--Matter Interactions, University of Southern Denmark, Campusvej 55, DK-5230 Odense M, Denmark}

\date{\today}

\begin{abstract}
\vspace{0.0cm}
\textbf{Abstract.} 
Electron-beam-induced luminescence typically relies on direct excitation by high-energy primary electrons. 
Here, we explore properties of an alternative excitation approach where cathodoluminescence (CL) is driven by substrate-generated electrons rather than by the primary electron beam.
Using color centers in diamond as sensitive and durable local probes, we investigate the spatial profiles of such indirect CL in different geometries and substrates. 
Photon-correlation experiments demonstrate increased synchronization of emitters at reduced currents, which we propose as a method for extracting the effective indirect excitation currents experienced by the emitters.
This approach enables the estimation of remarkably low currents, down to 0.1\,pA, highlighting the potential of substrate-assisted excitation for minimally invasive probing of sensitive emitters in CL microscopy. 
\vspace{0.3cm}
\end{abstract}

\maketitle

\section{Introduction}
Cathodoluminescence (CL) microscopy uniquely combines spectral, spatial, and temporal information, providing comprehensive insights into emission properties of quantum emitters, plasmonic resonators, and photonic nanostructures~\cite{Coenen2017,Polman2019,GarciadeAbajo2025}.
Fundamentally, all CL experiments depend on interactions between electrons and the sample.
These interactions may involve direct excitation, where the electron beam penetrates the sample and transfers energy directly to it. 
Alternatively, in aloof excitation, electrons do not directly penetrate the sample; instead, they pass very close to its surface, interacting with the evanescent near-field components of electromagnetic modes confined at the sample surface~\cite{Akerboom2024}.
A third, less-explored category is indirect excitation of emitters, which has been attributed to be mediated by secondary electrons (SEs)~\cite{Meuret2018,Mauser2021,Iyer2023} or backscattered electrons (BSEs) from the substrate~\cite{Narvez2013,Mauser2021}; however, the exact physical mechanisms involved and their impact on emitter excitation are not yet fully understood.
Notably, indirect excitation has been reported to induce significant photon bunching, observed in nitrogen vacancy (NV) color centers in diamond nanoparticles~\cite{Iyer2023}.

Photon-statistics measurements represent a powerful tool in CL microscopy, offering essential insights into excitation dynamics and quantum properties of emitters. 
Central to these measurements is the second-order autocorrelation function, 
$g_2(\tau)$, which characterizes temporal correlations between photon emission events separated by a delay time $\tau$. 
Analyzing $g_2(\tau)$ provides direct evidence of quantum optical phenomena such as photon antibunching or bunching~\cite{Tizei2013,Meuret2015,Fiedler2023}, enables extraction of emitter lifetime and excitation efficiency without requiring pulsed electron beams~\cite{SolGarcia2021,Fiedler2023}, and facilitates distinguishing between coherent and incoherent emission processes~\cite{Scheucher2022,Yanagimoto2025}.
Moreover, the amplitude of photon bunching $g_2(0)$ is highly sensitive to the electron beam current $I$, exhibiting an inverse proportionality [$g_2(0)\sim 1/I$]~\cite{Meuret2015,Feldman2018,Fiedler2023,FiedlerWS2}.

In this letter, we investigate the origin and spatial extent of indirect electron-beam excitation of quantum emitters. 
By employing silicon vacancy (SiV$^-$) color centers in diamond as local probes, we reveal the critical role of substrates in assisting indirect excitation. 
We examine how substrate atomic number ($Z$) and density ($\rho$) influence electron generation and spatial distribution in indirect excitation, uncovering the dominant role of BSEs. 
Furthermore, photon-correlation spectroscopy enables us to quantify the effective electron currents experienced by quantum emitters under indirect excitation, and its dependency on the emitter-substrate distance. 
Remarkably, we find that the effective excitation current can be reduced by several orders of magnitude compared to the electron-beam current used in the instrument, suggesting indirect excitation as a promising, low-damage approach for investigating sensitive quantum emitters in CL microscopy.

\begin{figure*}[t!]
\includegraphics[width=0.85\linewidth]{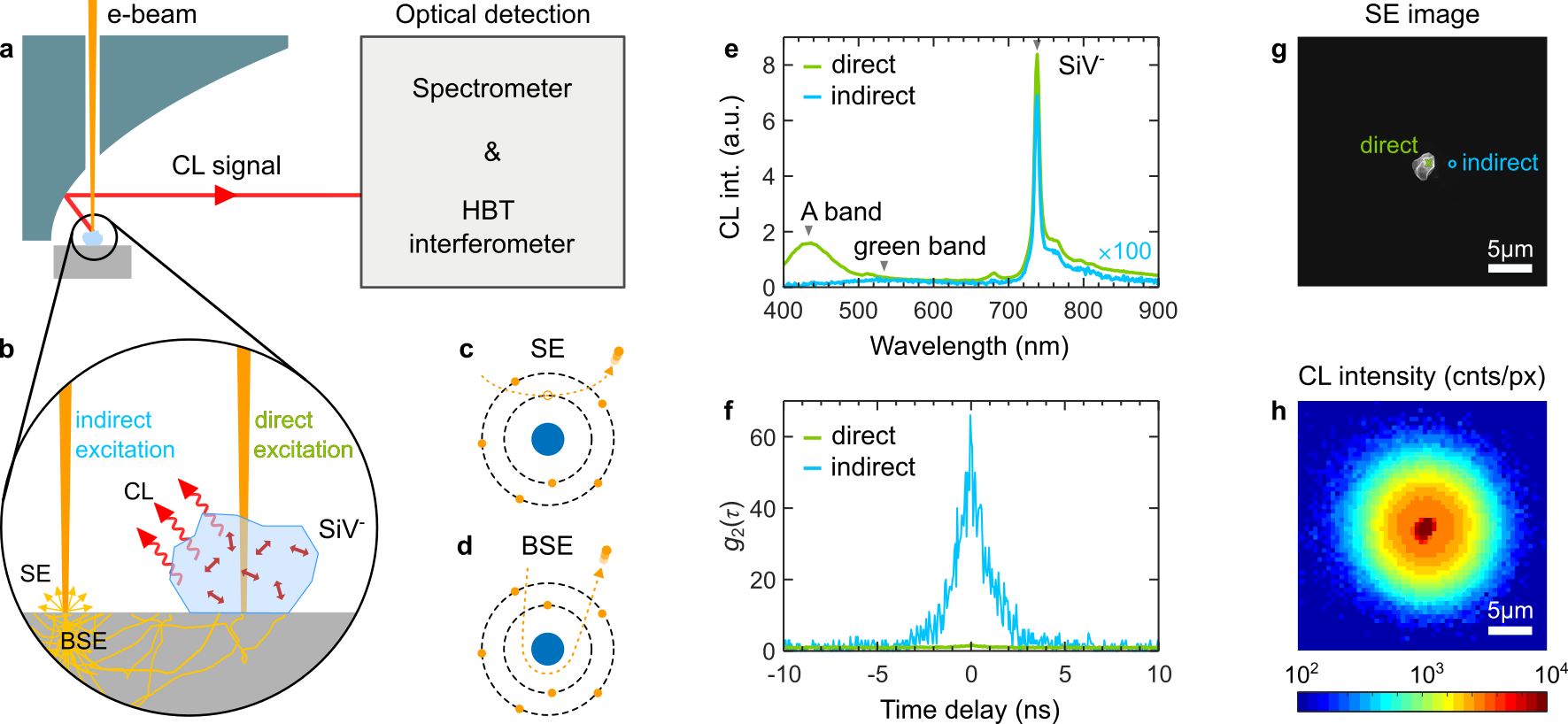}
\caption{\textbf{Experimental setup and electron-beam excitation pathways.}
\pnl{a} Schematic of the cathodoluminescence (CL) detection setup, utilizing a spectrometer and Hanbury~Brown--Twiss (HBT) interferometer. 
\pnl{b} Close-up illustration comparing indirect excitation (via secondary electrons (SE) and backscattered electrons (BSE) generated in substrate) and direct excitation of SiV$^-$ color centers in diamond.
 {\pnl{c}~Inelastic collisions between the incoming primary electrons and substrate atoms generate numerous low‑energy SEs.
\pnl{d}~An elastic collision deflects a primary electron back toward the surface with much of its initial energy retained, creating a high‑energy BSE that can escape the material.}
\pnl{e,f} CL spectra and photon-correlation measurements for direct (green) and indirect (blue) excitation at the same current, obtained from positions indicated in SE map in panel~\pnl{g} (scale bar is 5\,$\upmu$m).
\pnl{h} Corresponding to panel~\pnl{g} log‑scaled spatial map of SiV$^-$ CL, where the dark‑red core marks the intensity from directly excited centers in the diamond, and the concentric fall‑off captures the radially decaying indirect excitation of SiV$^-$ CL  {on a gold substrate} (pixel size $500\times500$\,nm$^2$, exposure time 0.2\,s). 
}
\label{fig-intro}
\end{figure*}

\section{Results and discussion}
\subsection{Sample and experimental setup}
We investigate electron-beam-driven CL excitation using diamond crystals containing SiV$^-$ color centers (Ad{\'a}mas Nanotechnologies)~\cite{Shames2020}. 
These diamonds are selected due to their high brightness, exceptional stability and robustness under electron beam exposure, showing no signs of degradation. 
Our experimental setup, schematically illustrated in Fig.~\ref{fig-intro}\pnl{a}, employs a scanning-electron microscope (SEM) equipped with a parabolic mirror positioned above the sample stage (TESCAN MIRA3). The mirror, featuring a central aperture for electron beam passage, collects generated CL signal and redirects it to the optical detection system~(SPARC Spectral, Delmic). 
This detection setup comprises a spectrometer for spectral analysis and a Hanbury~Brown--Twiss (HBT) interferometer for photon-correlation measurements, as detailed previously in our earlier works~\cite{Fiedler2023,FiedlerWS2}.

\subsection{Direct and indirect excitation}
Fig.~\ref{fig-intro}\pnl{b} conceptually illustrates the direct and indirect excitation mechanisms. 
Direct excitation occurs when the electron beam impinges on the diamond crystal, as indicated by the green cross in the SE image in Fig.~\ref{fig-intro}\pnl{g}. 
 {The energy of the incident primary electrons is predominantly transferred to the material through bulk plasmon excitation, whose decay into multiple electron–hole pairs subsequently drives the radiative emission of defects and color centers~\cite{Meuret2015}.}
The resulting CL spectrum in Fig.~\ref{fig-intro}\pnl{e} (green line) clearly displays emissions from the A-band, the green band, and a pronounced SiV$^-$ zero-phonon line (ZPL) at approximately 739\,nm~\cite{Zaitsev2001,Ebel2025}. 
The dominance of SiV$^-$ emission is due to the intentional doping, with typical crystals containing ensembles exceeding $10^3$ emitters  {(see Supplementary information (SI), Fig.~S1 and Ref. \cite{Shames2020})}. 
All presented spectra are corrected for instrument response, following the methodology outlined in our previous work~\cite{Ebel2025}. 
We select the CL signal of the SiV$^-$ line with optical bandpass filter at $(750\pm20)$\,nm (Thorlabs, FBH750-40) for analysis with an HBT setup.
This reveals that the direct excitation yields a modest photon-bunching peak with $g_2(0) \sim 1.6$~[green histogram in Fig.~\ref{fig-intro}\pnl{f}], consistent with the ensemble size and the relatively high electron-beam current $I$ of about 240\,pA~\cite{Fiedler2023}.

Remarkably, efficient CL of SiV$^-$ centers can also be excited indirectly. 
In this configuration, the 30\,keV electron beam, maintaining the same beam current of about 240\,pA, is positioned 2\,$\upmu$m away from the diamond crystal, as marked by the blue circle in Fig.~\ref{fig-intro}\pnl{g}. 
 {
Although the electron beam has a diameter of only about 10\,nm, ensuring that no part of it directly impinges on the diamond (see alignment verification in SI Section 2 and Fig.~S2), we still observe a CL signal from the SiV$^-$ centers (blue line in Fig.~\ref{fig-intro}\pnl{e}).
} 
The indirect excitation spectrum, exemplified by the blue line in Fig.~\ref{fig-intro}\pnl{e} (scaled by a factor of 100 for clarity), mimics the direct excitation spectrum with a prominent SiV$^-$ line at 739\,nm and the presence of the green band. 
Notably absent in indirect excitation is the A-band emission, likely due to its higher excitation threshold related to deep electronic transitions associated with dislocations or platelet defects~\cite{Zaitsev2001}.

 {To distinguish between direct, aloof, and indirect excitation, it is important to define their spatial ranges quantitatively. Excitation within approximately one beam diameter of the emitter corresponds to direct excitation, while aloof excitation arises from the evanescent electromagnetic field of the passing electron. The characteristic spatial extent of this aloof field is given by
$d_\mathrm{aloof}=v_\mathrm{el}\gamma_\mathrm{\epsilon}/\omega$, with $\gamma_\mathrm{\epsilon}=1/\sqrt{1-\epsilon v_\mathrm{el}^2/c^2}$, where $v_\mathrm{el}$ is the electron velocity, $\epsilon$ the dielectric permittivity, $c$ the speed of light, and $\omega$ the angular frequency of the excitation~\cite{RevModPhys.82.209}. 
For an electron energy of 30\,keV and an emitter wavelength of 739\,nm, we obtain $d_\mathrm{aloof}\approx 40$\,nm. Excitation observed at distances well beyond this scale can therefore be attributed to substrate-assisted indirect CL.
}

The spatial extent of indirect excitation is surprisingly large, as demonstrated in Fig.~\ref{fig-intro}\pnl{h}, which maps the SiV$^-$-related CL intensity surrounding the diamond shown in the corresponding SE image in Fig.~\ref{fig-intro}\pnl{g}. 
The spatial intensity distribution is extracted through spectral fitting methods (SI,  {Fig.~S3}). 
Photon-correlation measurements conducted under indirect excitation  {in Fig.~\ref{fig-intro}\pnl{f}} reveal a striking increase in photon bunching with $g_2(0) \sim 64$. 
The observed decrease in overall CL intensity by a factor of approximately 100 and the simultaneous thirty-fold increase in photon bunching strongly suggest a significant reduction in the effective excitation current. 
This inverse relationship between excitation current and photon bunching ($g_2 \propto 1/I$) has been observed in various solid-state luminescent systems~\cite{Meuret2015,Feldman2018,Fiedler2023,FiedlerWS2}.

To elucidate the underlying physical mechanism behind indirect excitation, we further investigate the role of substrate-generated electrons--namely, SEs and BSEs.
SEs are electrons ejected from the substrate material due to inelastic collisions with primary electrons [Fig.~\ref{fig-intro}\pnl{c}], typically possessing energies below 50\,eV~\cite{Seiler1983}. 
The schematics in Fig.~\ref{fig-intro}\pnl{b} illustrates the generation and escape trajectory of SEs, indicating their limited range and high density near the electron beam impact point. 
Conversely, BSEs result from elastic scattering events and retain a significant fraction of their initial kinetic energy  {[Fig.~\ref{fig-intro}\pnl{d}]}, typically several keV~\cite{Matsukawa1974}. 
These electrons, as shown in the schematics in Fig.~\ref{fig-intro}\pnl{b}, exhibit more extensive trajectories and can escape from deeper regions.

\begin{figure}[h]
\includegraphics[width=0.58\linewidth]{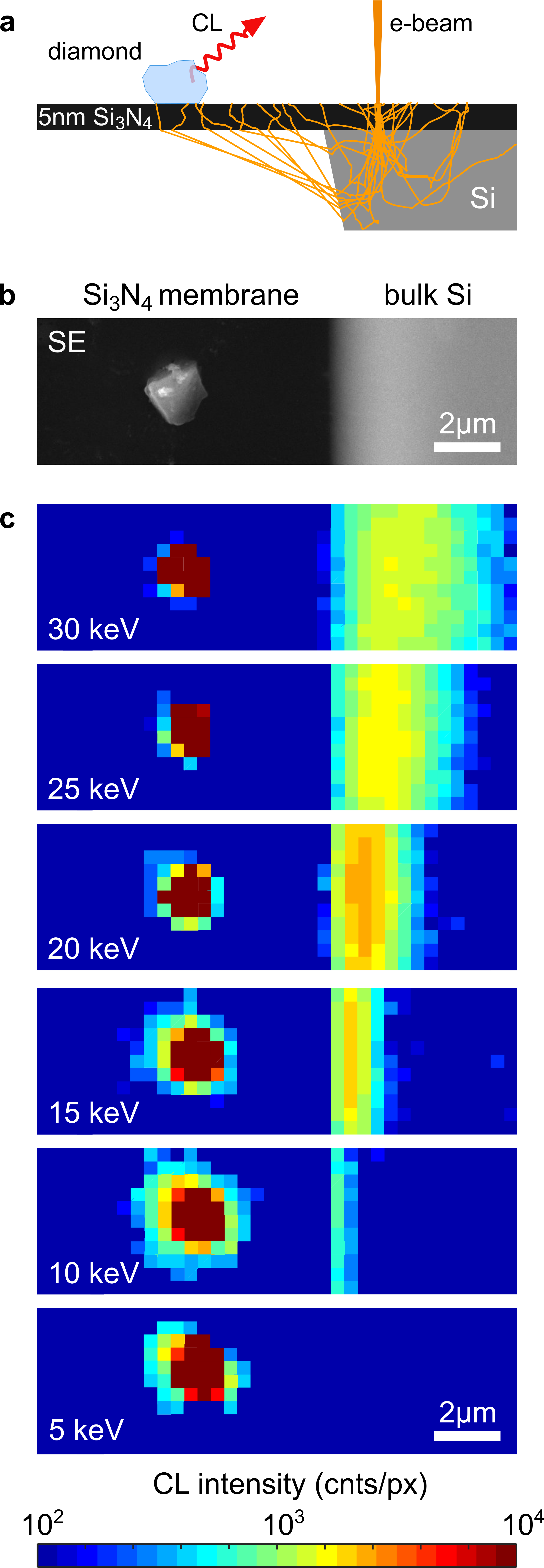}
\caption{\textbf{Indirect excitation of SiV$^-$ centers in a diamond at the edge of a thin $\mathbf{\mathrm{Si}_3\mathrm{N}_4}$ membrane.}
\pnl{a}~Schematic of the experimental geometry: an electron beam impinges on the bulk part of substrate, generating BSEs (orange trajectories) that excite CL of SiV centers in a diamond placed on a 5\,nm $\mathrm{Si}_3\mathrm{N}_4$ membrane.
\pnl{b}~SE image of the sample, showing the diamond on the $\mathrm{Si}_3\mathrm{N}_4$ membrane region (left) and the bulk Si support frame (right).
\pnl{c}~CL intensity maps recorded at beam energies from 30\,keV to 5\,keV. 
Note the the logarithmic colormap scaling (pixel size $500\times500$\,nm$^2$, exposure time 1\,s). 
}
\label{fig-membrane}
\end{figure}

\subsection{Role of substrate}
Substrate composition and electron beam energy are critical parameters for indirect excitation, as they directly influence the generation and spatial range of SEs and BSEs.
Fig.~\ref{fig-membrane}\pnl{a} illustrates the experimental configuration utilized to investigate the substrate's role. Diamond crystals containing SiV$^-$ centers are dropcast onto ultra-thin (5\,nm) silicon nitride ($\mathrm{Si}_3\mathrm{N}_4$) membrane windows supported by bulk silicon (Si) frames. This configuration allows us to independently assess the indirect excitation triggered by electrons interacting either with the thin $\mathrm{Si}_3\mathrm{N}_4$ membrane or the underlying Si frame substrate.

In Fig.~\ref{fig-membrane}\pnl{b}, we show a representative SE image of one such diamond crystal located approximately 3\,$\upmu$m from the edge of the Si supporting frame. To systematically assess indirect excitation, we performed electron-beam scans over the $\mathrm{Si}_3\mathrm{N}_4$ membrane and the adjacent bulk Si frame at various electron beam energies ranging from 5\,keV to 30\,keV, keeping the beam current constant at about 1400\,pA. The resulting CL intensity maps are presented in Fig.~\ref{fig-membrane}\pnl{c}.
Analyzing these CL maps, we first note a critical observation: at the highest electron beam energy (30\,keV), SiV$^-$ centers in the diamond crystal show negligible excitation when the electron beam is positioned on the $\mathrm{Si}_3\mathrm{N}_4$ membrane alone. 
However, when the beam impacts the adjacent bulk Si supporting frame, significant SiV$^-$ CL intensity is clearly detected several micrometers away. 
This highlights an essential finding--the presence of a substantial substrate is necessary for efficient indirect excitation.

Upon reducing the electron beam energy from 30\,keV down to 5\,keV in Fig.~\ref{fig-membrane}\pnl{c}, two trends become immediately apparent. 
Firstly, the spatial range of the indirect excitation by Si frame gradually shortens as the beam energy decreases, becoming negligible at 5\,keV. 
Secondly, when the beam is on $\mathrm{Si}_3\mathrm{N}_4$ membrane, CL intensity halo appears around the diamond at low beam energies.
To explain these trends, we performed Monte Carlo simulations of the BSE spatial behavior with respect to electron-beam energy (see SI, Section~2)~\cite{Drouin2007}.
For bulk Si frame, the BSE yield varies only weakly over the 5–30\,keV range [SI,  {Fig.~S4\pnl{a}}], whereas the diffusion range of BSEs significantly increases with electron-beam energy [SI,  {Fig.~S4\pnl{b}}].
Therefore, high-energy beams generate BSEs that can travel micrometres and excite the diamond from the distant Si frame, while at 5\,keV their reach is limited.
Conversely, lowering the beam energy increases the probability that the electron beam interacts with the $\mathrm{Si}_3\mathrm{N}_4$ membrane, which raises the local BSE yield but with a shorter diffusion range, producing the observed low‑energy CL brightening around the diamond in Fig.~\ref{fig-membrane}\pnl{c}.

 {
These results indicate that indirect excitation is predominantly driven by BSEs, with only a minor contribution from substrate-generated SEs. Although approximately 40\% of emitted SEs are generated by BSEs~\cite{Kanter1960}, implying that BSE spatial distributions are inherently reflected in SE signals~\cite{Joy1987,Seiler1983}, bias-controlled CL mapping experiments show that suppressing SE emission does not significantly alter the spatial excitation profile (see SI, Fig.~S5). This demonstrates that SEs do not play a dominant role in indirect excitation. Consequently, we attribute the observed micrometer-range indirect excitation to BSEs, while low-energy SEs ($<50$\,eV), characterized by nanometer-scale escape depths, are unlikely to account for the long-range excitation observed here~\cite{Kanter1960,Reimer1977}.
}

\subsection{Effect of substrate composition}
To further probe the role of BSEs in indirect excitation of CL, we tested the response of different substrates by varying their atomic number $Z$ and density $\rho$. Since BSE generation arises primarily from elastic scattering of incident electrons with atomic nuclei, both the yield and angular distribution of BSEs are strongly dependent on the substrate composition~\cite{Jost1963,Yasuda1995}.
High-$Z$ and high-$\rho$ materials produce larger BSE yields and shorter diffusion range, whereas low-$Z$ and low-$\rho$ materials allow deeper penetration and broader BSE escape profiles.

\begin{figure}[t]
\includegraphics[width=0.99\linewidth]{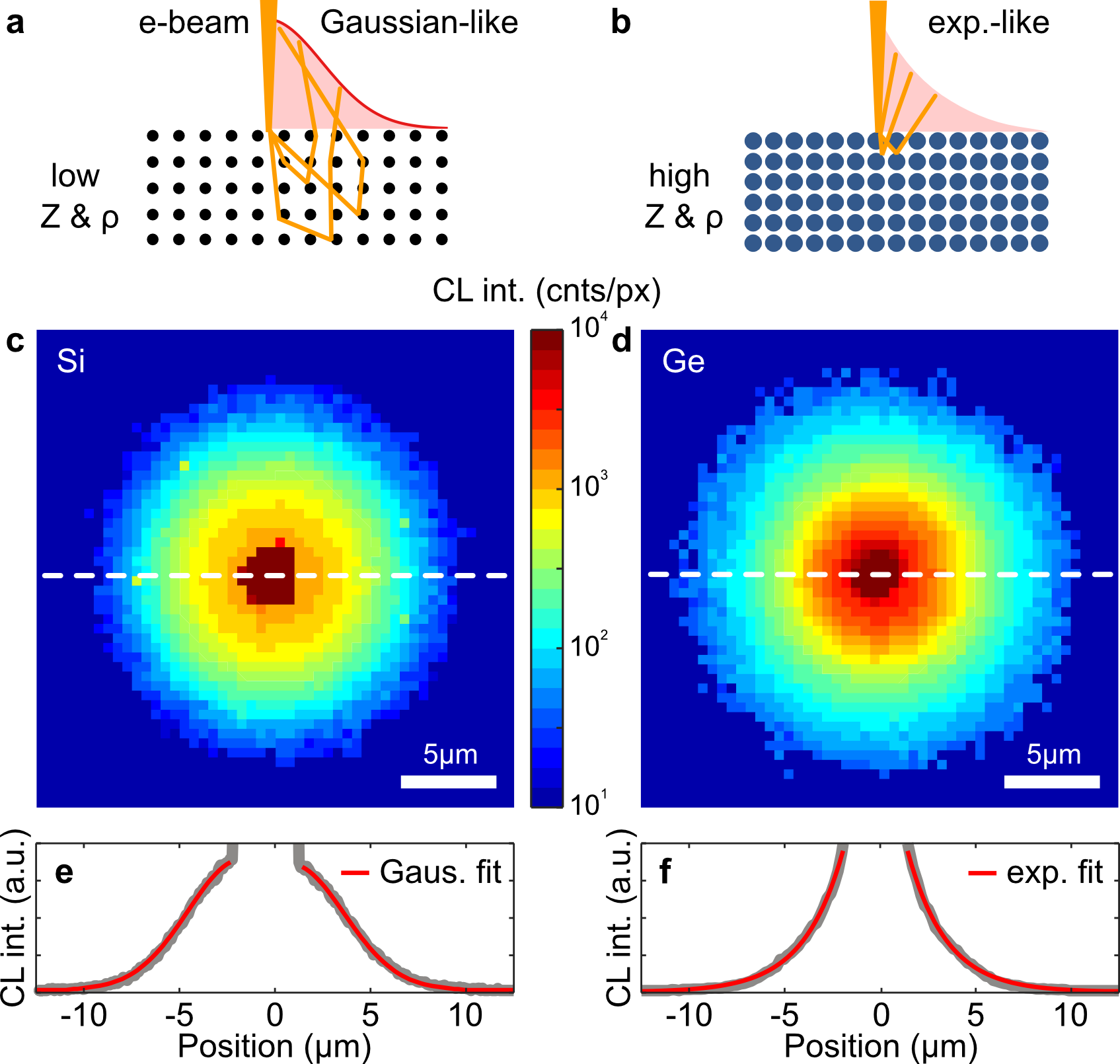}
\caption{\textbf{Influence of substrate material properties on indirect excitation.}
\pnl{a} and \pnl{b} schematically illustrate the BSE trajectories (orange lines; from Monte Carlo simulations~\cite{Drouin2007}) and resulting intensity distributions for substrates with low atomic number ($Z$) and density ($\rho$), such as Si, and substrates with high $Z$ and density, such as Ge. In low-$Z$ and low-density substrates \pnl{a}, BSEs originate from deeper within the material, yielding a Gaussian-like intensity distribution.
Conversely, in high-$Z$ and high-density substrates \pnl{b}, BSE trajectories originate predominantly from shallow regions, resulting in an exponential-like intensity distribution. 
Panels~\pnl{c,d} present experimental CL intensity maps for indirect excitation with 30\,keV and 1.4\,nA on silicon (Si) and germanium (Ge) substrates, respectively. 
Note the logarithmic colormap scaling [pixel size $500\times500$\,nm$^2$, exposure time 5\,ms in \pnl{c} and 50\,ms in \pnl{d}].
Horizontal dashed lines indicate the cross-sectional locations for extracting intensity profiles shown in panels~\pnl{e,f}. 
These intensity profiles are accurately described by Gaussian (for Si) and exponential (for Ge) fitting functions (red lines).
}
\label{fig-bulk}
\end{figure}

We conducted comparative experiments on bulk Si and Ge substrates, which differ considerably in atomic number and density. 
The expected distributions for electron backscattering patterns in these substrates are schematically presented in Fig.~\ref{fig-bulk}\pnl{a,b}. 
In lower-density and lower-$Z$ substrates, such as Si ($Z=14$, $\rho\simeq 2.3$\,g/cm$^3$), electrons penetrate deeper, undergoing multiple elastic scattering events before being emitted from a broader region. 
This multiple scattering process generally leads to a Gaussian-like spatial distribution~\cite{Jost1963,Yasuda1995}. 
Conversely, substrates with higher atomic number and density, such as Ge ($Z=32$, $\rho\simeq 5.3$\,g/cm$^3$), result in electrons being scattered closer to the surface, producing an exponential-like spatial distribution dominated by fewer, more localized scattering events~\cite{Yasuda1995,Murata1974}.

Experimentally validating these expectations, we deposited diamond crystals containing SiV$^-$ centers on Si and Ge substrates and performed electron-beam-induced indirect excitation experiments using a constant electron-beam current of approximately 1400\,pA and an acceleration voltage of 30\,keV. 
Fig.~\ref{fig-bulk}\pnl{c,d} depict the experimentally acquired CL intensity maps for indirect excitation mediated by the Si and Ge substrates, respectively. These intensity maps indeed display pronounced differences, which we demonstrate on cross-sectional intensity profiles in Fig.~\ref{fig-bulk}\pnl{e,f}.
These profiles are extracted along the horizontal dashed lines indicated in Fig.~\ref{fig-bulk}\pnl{c,d} and reveal the expected differences in spatial emission profiles.
Specifically, the CL intensity profile on the low-Z and $\rho$ Si substrate follows a Gaussian function [red line in Fig.~\ref{fig-bulk}\pnl{e}], characteristic of broader electron emission areas arising from multiple scattering events. 
In contrast, the profile on the high-Z and $\rho$ Ge substrate closely matches an exponential decay function [red line in Fig.~\ref{fig-bulk}\pnl{f}], consistent with electrons originating from fewer, shallow scattering events near the substrate surface.  {From these fits, we obtain for the Si substrate a characteristic Gaussian width of $\sigma_\mathrm{Si}=3.11\pm0.50,\mu\mathrm{m}$, while for the Ge substrate we extract an exponential decay length of $L_\mathrm{Ge}=1.84\pm0.50,\mu\mathrm{m}$. The quoted uncertainties correspond to the quadratic sum of the statistical fitting uncertainty and a systematic contribution arising from the finite pixel size of the scan.}

\begin{figure*}[ht!]
\includegraphics[width=0.8\linewidth]{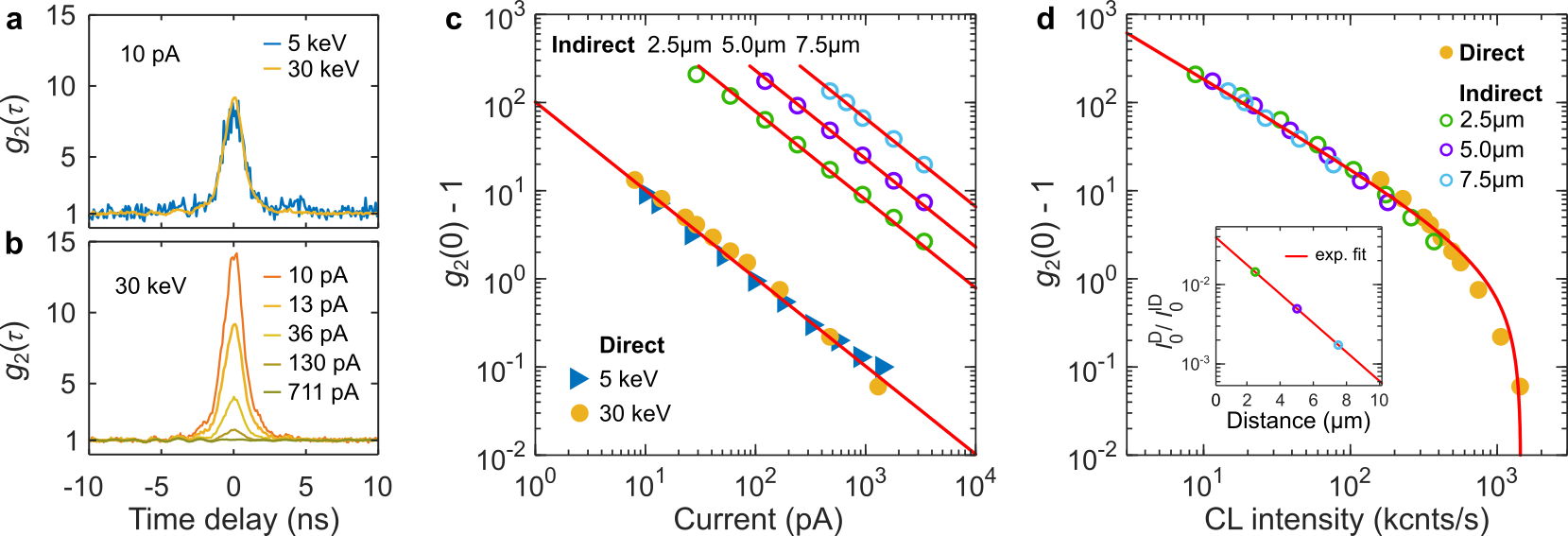}
\caption{\textbf{Photon bunching in indirect excitation.} All data is shown for the same diamond crystal on gold substrate. 
\pnl{a}~Photon correlation $g_2(\tau)$ at 10\,pA for direct excitation shows no dependence on electron beam energy (5\,keV vs. 30\,keV).
\pnl{b}~Photon correlation $g_2(\tau)$ at 30\,keV for direct excitation demonstrates strong dependence on electron-beam current.
\pnl{c}~Photon bunching amplitude as a function of current for direct and indirect excitation at various distances.
\pnl{d}~Photon bunching amplitude plotted versus CL intensity for indirect and direct excitation. 
Inset shows the exponential decay of excitation efficiency of indirect CL with distance.
}
\label{fig-bunching}
\end{figure*}

Additionally, we examined diamond crystals placed on graphite ($Z=6$, $\rho\simeq 2.3$\,g/cm$^3$) and gold ($Z=79$, $\rho\simeq 19.3$\,g/cm$^3$) substrates (SI,  {Fig.~S6}). 
These complementary experiments yielded analogous results, with graphite substrate producing Gaussian-like distributions and gold substrate exhibiting exponential-like spatial profiles.
Furthermore, we measured the CL spatial profiles of a diamond on Si substrate at different electron beam energies, and observed a transition from Gaussian-like distribution at 30\,keV to exponential-like one at 5\,keV, consistent with the expected energy-dependent behavior of BSE scattering (SI,  {Fig.~S4 and Fig.~S7}).
The observation of indirect excitation in a light-element substrate such as Si at a low electron beam energy of 5\,keV also excludes x-ray generation as the primary excitation mechanism.
 {
Monte Carlo simulations using CASINO~v2.42 reproduce the qualitative substrate- and energy-dependent trends of indirect excitation (see SI, Section 4) but underestimate the measured excitation range, preventing a fully quantitative comparison due to variations in diamond geometry and emitter distribution.
}

\subsection{Effective current}
Next, it is essential to quantify the effective electron current experienced by an SiV$^-$ ensemble under indirect excitation as it cannot be measured directly by the instrument.
For this, we propose using photon-correlation measurements, which exhibit a strong and well-defined dependence on excitation current. 
Unlike conventional photoluminescence, electron-beam-induced CL often produces a pronounced photon bunching peak at zero time delay, which is highly sensitive to variations in the electron-beam current~\cite{Meuret2015,Feldman2018,Fiedler2023,FiedlerWS2}. 
First, we demonstrate how photon bunching responds to electron-beam parameters under direct excitation.
Fig.~\ref{fig-bunching}\pnl{a} shows photon-correlation histograms of SiV$^-$ ensemble in a diamond on gold substrate, which were acquired at the same electron-beam current of 10\,pA but for two acceleration voltages, 5\,keV and 30\,keV.
While the bunching amplitude remains unaffected by the acceleration voltage in Fig.~\ref{fig-bunching}\pnl{a}, it is highly sensitive to the electron-beam current in Fig.~\ref{fig-bunching}\pnl{b}, where higher currents lead to reduced bunching amplitudes.
Fig.~\ref{fig-bunching}\pnl{c} summarizes these observations, displaying the photon bunching magnitude as a function of electron-beam current for both 5\,keV (filled blue symbols) and 30\,keV (filled yellow symbols) under direct excitation. 
As expected, the photon bunching magnitude clearly exhibits an inverse relationship [$g_2(0) \sim 1/I$] with the beam current, as indicated by the red lines in Fig.~\ref{fig-bunching}\pnl{c}.
The overlapping curves for 5\,keV and 30\,keV further support the independence of photon bunching from acceleration voltage, while they demonstrate the photon bunching sensitivity to changes in electron beam current. 

In order to quantify the effective current driving indirect excitation, we positioned the electron beam at various distances (2.5\,$\upmu$m, 5\,$\upmu$m, and 7.5\,$\upmu$m) from the diamond emitter and measured photon-correlation histograms. 
Fig.~\ref{fig-bunching}\pnl{c} presents these results, showing again an inverse relationship with applied current by instrument. 
However, we observe an upward shift of the photon bunching curves toward higher $g_2(0)$ values for indirect excitation positions. 
This shift clearly indicates a reduction in effective current at the emitter location due to indirect excitation conditions.
Moreover, the observation of photon bunching confirms the electron-based origin of indirect excitation, effectively excluding the possibility of excitation by substrate-generated photons.

Simultaneously, we acquired the CL intensity of the SiV$^-$ emission line at these indirect excitation positions, allowing us to plot the photon bunching amplitude versus CL intensity in Fig.~\ref{fig-bunching}\pnl{d}. 
Remarkably, data points for both direct and indirect excitation conditions lie along the same characteristic fitting curve (red). 
This curve describes the excitation efficiency relationship between photon bunching amplitude and emission intensity, as reported in our previous work~\cite{FiedlerWS2}  {and SI, Section 7}. 
The unified curve indicates that indirect excitation fundamentally shares the same excitation mechanism as direct excitation but operates at effectively reduced electron-beam currents.

Therefore, the photon bunching data can provide a robust method for determining the effective excitation current in indirect excitation scenarios. 
 {To extract numerical values for this effective current, we fit the measured photon bunching data in Fig.~\ref{fig-bunching}\pnl{c} with the inverse-current dependence 
\begin{equation}
    g_2(0)-1=\frac{I_0}{I},
    \label{eq_current}
\end{equation}
where $I_0$ represents the characteristic current required to have one electron exciting one bulk plasmon per lifetime~\cite{Meuret2015}. 
From fitting the data to Eq.~\ref{eq_current}, we obtain characteristic currents, $I_{0}^\mathrm{D}$ and $I_{0}^\mathrm{ID}$, for direct (D) and indirect (ID) excitation at each distance.} 
The ratio $I_{0}^\mathrm{D}/I_{0}^\mathrm{ID}$ represents the factor by which the effective excitation current $I_\mathrm{eff}$ is reduced compared to the nominal electron beam current $I$, directly quantifying the current driving  indirect CL. 
The inset of Fig.~\ref{fig-bunching}\pnl{d} summarizes these ratios, revealing an exponential decay of effective current with increasing distance (red curve), consistent with our previous findings of exponential scaling of CL intensity on high-$Z$ substrates (see Fig.~\ref{fig-bulk}).
Using photon-bunching measurements, we estimate that the effective excitation current experienced by the emitter under indirect excitation conditions is reduced by few orders of magnitude, below 0.1\,pA in the presented experiment.
 {
The dominant uncertainties in this estimate arise from systematic effects, primarily slow beam drift during the long acquisition times required for 
$g_2(\tau)$ measurements, which leads to uncertainty in the effective beam–emitter distance (SI, Fig.~S2). Additional uncertainty stems from the probe geometry, as non-uniform SiV distributions and asymmetric diamond shapes result in a non-uniform response to the spatially extended indirect excitation.
}

To further validate these observations, we repeated photon bunching experiments on diamonds placed on Si and Ge substrates (SI,  {Fig.~S8}), which provided direct support for the Gaussian and exponential scaling observed earlier. 
These additional measurements confirm that the scaling of effective current closely parallels that of CL intensity~\cite{FiedlerWS2}, as both are fundamentally governed by the same physical processes of electron-substrate interactions and subsequent BSE generation.

Our results collectively demonstrate that photon-bunching measurements in CL provide a precise and quantitative tool for assessing the effective electron current in indirect electron-beam excitation, offering new capabilities for controlling and optimizing electron-induced luminescence in nanoscale emitters.
 {Direct and indirect excitation are governed by the same plasmon-mediated energy-transfer mechanism and therefore lead to identical excitation dynamics. Indirect excitation mainly reduces the effective excitation current and alters the spatial excitation profile, without affecting the intrinsic radiative recombination efficiency. An absolute CL quantum efficiency cannot be uniquely defined, since a single electron can generate multiple plasmons and emitted photons, although photon-statistics–based approaches may enable such estimates in the future~\cite{Varkentina2022,Yanagimoto2025}.}

\section{Conclusions}
In conclusion, we studied indirect electron-beam excitation of SiV$^-$ centers in diamond showing that it is dominantly driven by BSEs generated in nearby substrates. 
We found that the spatial distribution of indirect excitation depends strongly on substrate atomic number $Z$ and density $\rho$, with Gaussian-like distributions observed for low-$Z$ and low-$\rho$ materials and exponential-like for high-$Z$ and high-$\rho$ materials. 
Photon-correlation measurements revealed that indirect excitation reduces the effective excitation current experienced by emitters below $\sim$0.1\,pA, highlighting indirect CL as a gentle, low-damage, and spatially tunable nanoscale excitation strategy for quantum emitters.\\

\noindent
\textbf{Competing interests:}
 {The authors declare no competing interests.}

\noindent
\textbf{Funding Declaration:} The Center for Polariton-driven Light-Matter Interactions (POLIMA) is sponsored by the Danish National Research Foundation (Project No.~DNRF165).

\noindent
\textbf{Author contributions:} The idea of the project was conceived by S.~E. and S.~M. 
CL spectroscopy was carried out by S.~E. and S.~M. 
The project was supervised by S.~M. and N.~A.~M.
All authors contributed to analyzing the data and writing the manuscript. 
All authors have accepted responsibility for the entire content of this manuscript and approved its submission.

\noindent
 {\textbf{Data availability:} The data that support the findings of this study are available from the corresponding author upon reasonable request.}

\noindent
 {\textbf{Supplementary information:} The online version contains
supplementary material available at .}

\bibliography{bibliography}
\bibliographystyle{unsrt}

\clearpage
\newpage

\begin{figure*}[ht!]
\includegraphics[width=8.25cm]{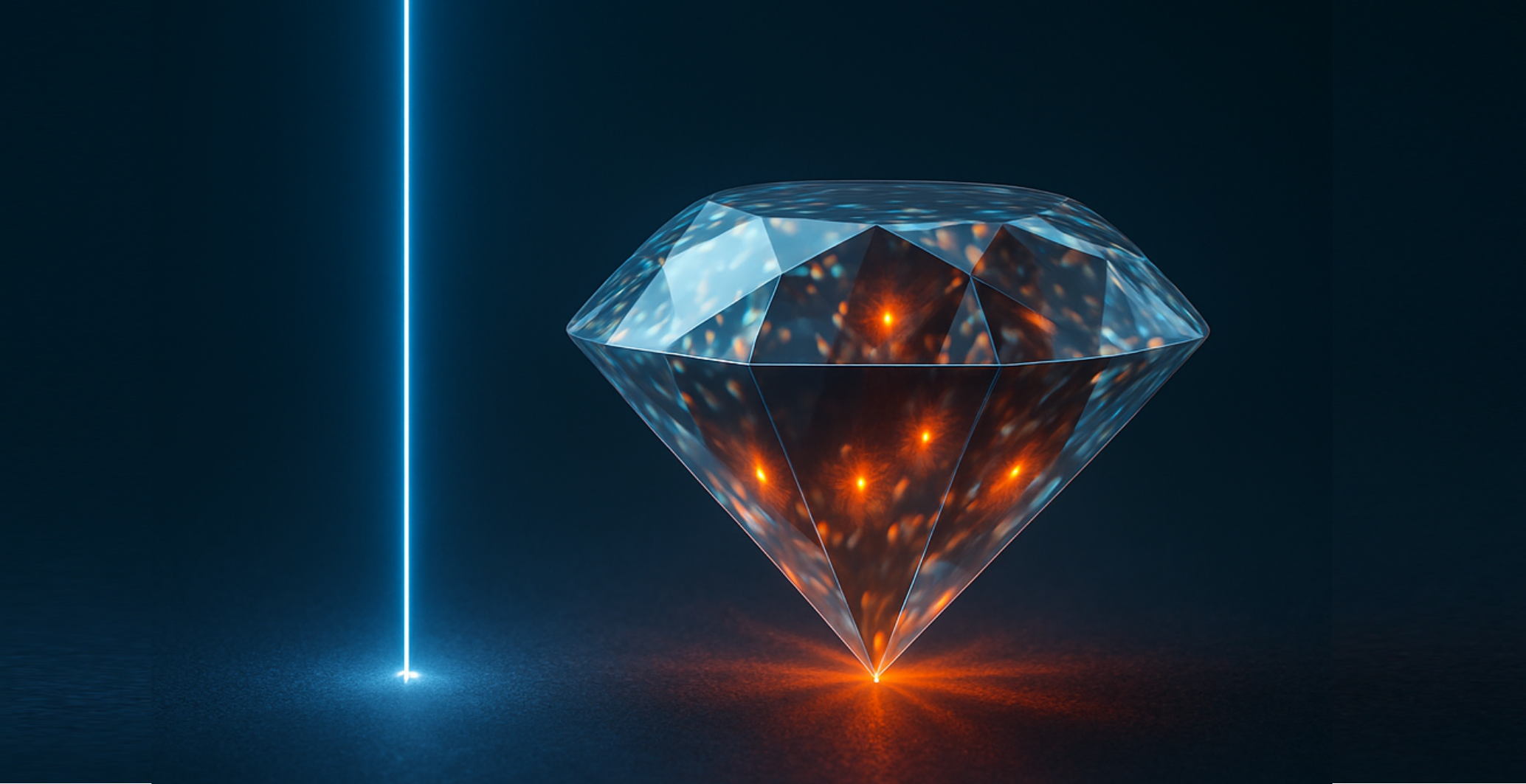}
\caption*{\textbf{Abstract Graphic} 
}
\end{figure*}

\end{document}